\documentclass[%
reprint,
superscriptaddress,
nofootinbib,
amsmath,amssymb,
pra,
]{revtex4-1}

\usepackage{graphicx,overpic}
\usepackage{dcolumn}
\usepackage{bm}

\usepackage{amssymb}
\usepackage{amsmath}
\usepackage{amsfonts}
\usepackage{xspace}
\usepackage{bm,bbm}
\usepackage{relsize}
\usepackage{siunitx}
\usepackage[usenames,dvipsnames]{color}

\renewcommand{\vec}[1]{\ensuremath{\bm{#1}}}
\newcommand{\bvec}[1]{\ensuremath{\bar{\bm{#1}}}}
\newcommand{\ie}{\emph{i.e.}\xspace}
\newcommand{\cf}{\emph{cf.}\xspace}

\newcommand{\identity}{\ensuremath{\mathlarger{\mathbbm{1}}}}
\renewcommand{\eqref}[1]{Eq.~\ref{eq:#1}}
\newcommand{\figref}[1]{Fig.~\ref{fig:#1}}
\newcommand{\tabref}[1]{Tab.~\ref{tab:#1}}
\newcommand*{\titlestyle}[1]{{({\textsf{\MakeLowercase{#1}}})}}

\newcommand*{\cc}[1]{\ensuremath{\,\overline{#1\vphantom{\bar{#1}}}\,}}
\newcommand{\wolog}{w.l.o.g.\xspace}

\newcommand{\imag}[1]{\text{Im}\left\{#1\right\}}

\begin{document}

\title{A group theoretical route to deterministic Weyl points in chiral photonic lattices}

\author{Matthias Saba}
\email[Email: ]{m.saba@imperial.ac.uk}
\author{Joachim M.~Hamm}
\affiliation{The Blackett Laboratory, Imperial College London, London SW7 2AZ, UK}
\author{Jeremy J.~Baumberg}
\affiliation{The Cavendish Laboratory, University of Cambridge, Cambridge CB3 0HE, UK}
\author{Ortwin Hess}
\email[Email: ]{o.hess@imperial.ac.uk}
\affiliation{The Blackett Laboratory, Imperial College London, London SW7 2AZ, UK}

\date{\today}

\begin{abstract}
Classical topological phases derived from point degeneracies in photonic bandstructures show intriguing and unique behaviour. Previously identified exceptional points are based on accidental degeneracies and subject to engineering on a case-by-case basis. Here we show that symmetry induced (deterministic) pseudo Weyl points with non-trivial topology and hyper-conic dispersion exist at the centre of the Brillouin zone of chiral cubic systems. We establish the physical implications by means of a $P2_13$ sphere packing, realised as a nano plasmonic system and a photonic crystal.
\end{abstract}


\maketitle

Current broad interest in topological phases, triggered by the discovery of the quantum Hall effect \cite{PhysRevLett.45.494} and its theoretical investigation \cite{PhysRevLett.49.405,PhysRevB.48.11851,KOHMOTO1985343}, can mainly be attributed to the fact that topological features are, due to their discrete nature, insensitive to system perturbations, and can, for example, give rise to the existence of topologically induced edge states for bulk systems \cite{RevModPhys.83.1057,RevModPhys.82.3045}. Plasmonic \cite{PhysRevB.93.241402} and single electron \cite{Siroki2016} surface states of Weyl semi-metals, with an isolated exceptional point of non-trivial topology, are stable against perturbations and give rise to peculiar dynamics. Recently, it has been demonstrated that topological quantization occurs in entirely classical systems such as two-dimensional (2D) photonic crystals \cite{PhysRevLett.93.083901,PhysRevA.78.033834}, sparking a new wave of research on photonic topology \cite{Lu2014}. In particular, topologically protected Weyl points with hyperconic dispersion have been identified in double gyroid photonic crystals with broken parity-time symmetry \cite{Lu2013}.
Concurrently, group theory provides a tool to predict whether a given spatio-temporal symmetry permits topologically non-trivial exceptional points, or induces them deterministically. This idea has successfully found its way and been applied to classical \cite{PhysRevB.89.134302} and quantum mechanical \cite{PhysRevLett.116.186402,Bradlynaaf5037} systems. Indeed, the existence of deterministic two and three-fold degeneracies at the center of the Brillouin zone (BZ), aka the $\Gamma$ point, for cubic symmetries is well known and documented in the literature \cite{Bradley_GT}. Recently, it has been shown that some of these degeneracies are topologically non-trivial in electronic systems \cite{1957JPCS1249K,Bradlynaaf5037}.
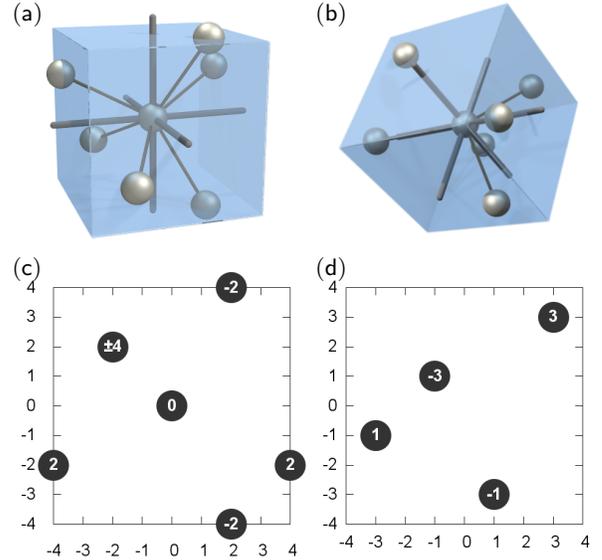
\begin{figure}[t]
    \begin{overpic}[width=.45\textwidth]{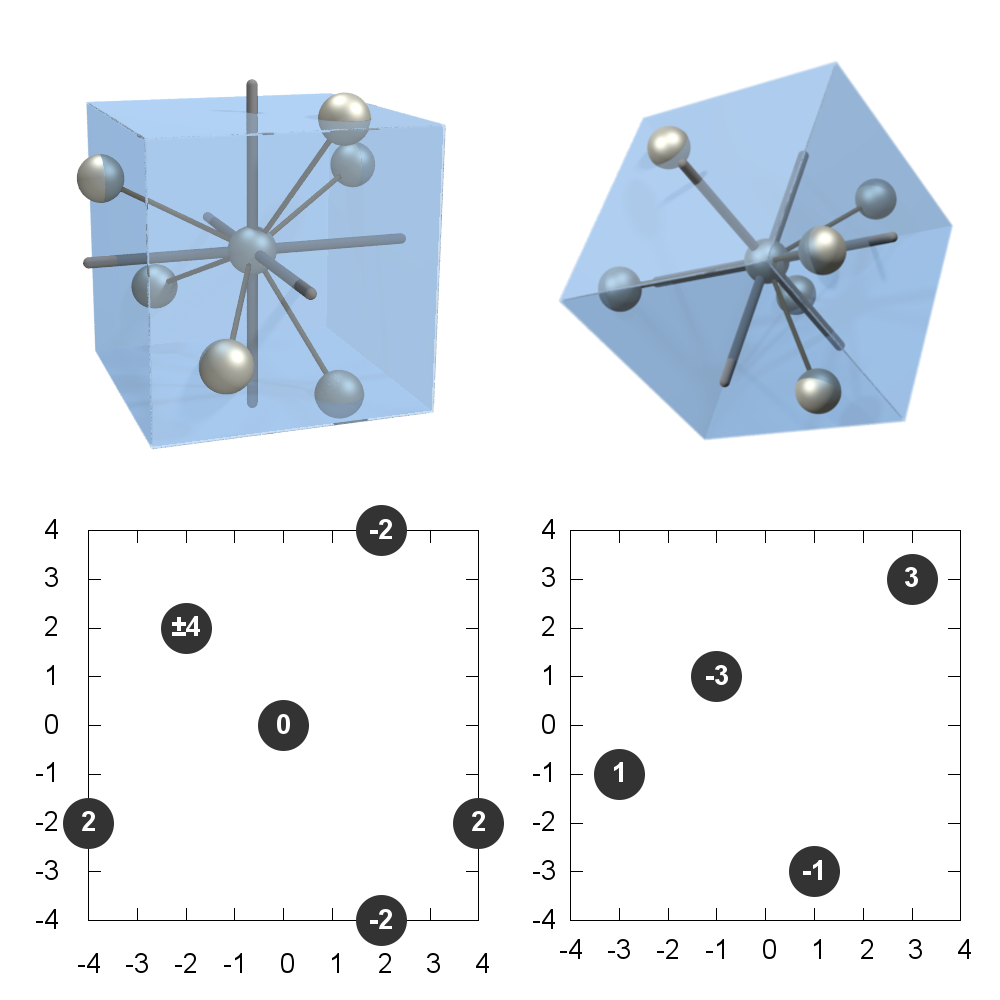}
        \put(2,91){\titlestyle{A}}
        \put(52,91){\titlestyle{B}}
        \put(2,49){\titlestyle{C}}
        \put(52,49){\titlestyle{D}}
    \end{overpic}
    \vspace{-1em}
    \caption{(Color online) Illustration of the $P2_13$ sphere packing. \titlestyle{A} The glass cube shows the simple cubic unit cell, that is centred at the position of one of the spheres, whose $6$ nearest neighbours lie on the cubes' facets. The thick Cartesian rods, and the thin connection rods are shown for illustration purposes only. \titlestyle{B} The same cube shown from the $[111]$ direction. \titlestyle{C} Projection of \titlestyle{A} onto the $[001]$ plane, with spatial unit $a/8$ and $z$ coordinate in the respective sphere. \titlestyle{D} Same as \titlestyle{C}, but with crystallographic choice of origin \cite{0792365909}.}
    \label{fig:spheres}
\end{figure}

Here we show on the basis of group and perturbation theory that symmetry induced three-fold degenerate \emph{pseudo Weyl points} (PWPs) exist at the $\Gamma$ point in classical (photonic) systems. They split isotropically in first order in the Bloch wave vector $\vec{k}$ for any chiral cubic space group with time reversal symmetry. The PWPs studied here constitute a deterministic 3D analog to previously studied accidental Dirac points \cite{Huang2011}. We show and demonstrate that they are of non-trivial topology, leading to protected surface states. In this letter, we first derive a 3D perturbation model that leads to hyperconic dispersion with non-trivial topology, and an intermediate flat band. We then construct a minimalistic geometry, a $P2_13$ sphere packing (\figref{spheres}), which satisfies the symmetry requirements, and apply it to a quasistatic coupled-dipole model, before discussing topologically protected surface states that emanate from a PWP in a photonic crystal analog. This underscores that the existence of PWPs, including the peculiar transport properties of associated bulk and surface states, only depends on the underlying symmetry irrespective of the particular physical realization. 

The theory applies to all linear and self-consistent physical systems with time reversal invariance and chiral cubic symmetry, with dynamics described by a Fourier integral over Hilbert states $|v(\omega)\rangle\in\mathcal{H}$ which individually solve a homogeneous (generally non-linear) frequency domain eigenproblem $M(\omega)\,|v(\omega)\rangle = 0$. Symmetry requires that the $\mathcal{H}$ operator $M$ commutes with all elements of the underlying space group $\mathcal{G}$, represented by $\mathcal{H}$ operators $g$. As a consequence, a set of $N$ degenerate eigenvectors $|v_n(\omega)\rangle$ form a \emph{representation} of $\mathcal{G}$, i.e.~they span an $N$-dimensional vector space with an associated algebra that is homomorphic to $\mathcal{G}$. \emph{Irreducible representations} impose a lower limit on the dimensionality of the respective vector space, resulting in deterministic degeneracies \cite{Bradley_GT,Dresselhaus,PhysRevB.88.245116}.
\begin{figure*}[t]
    \centering
    \begin{overpic}[width=1.\textwidth]{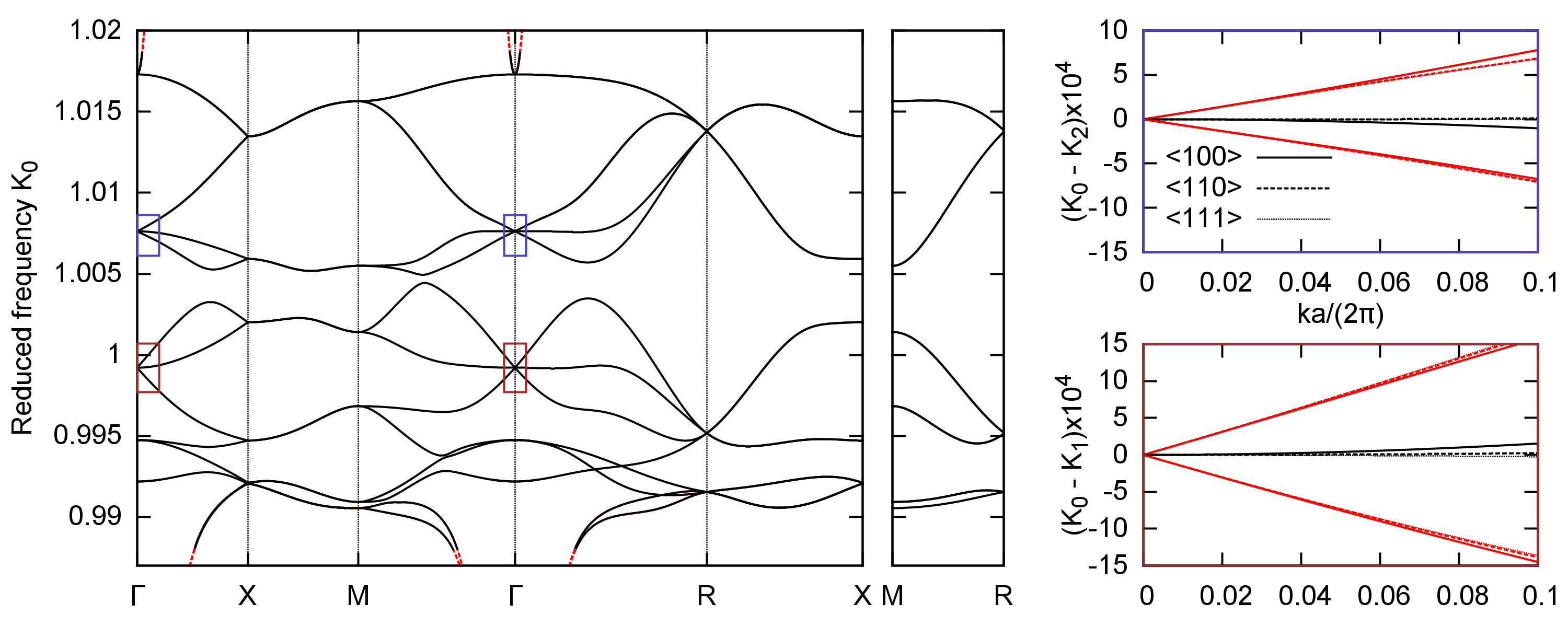}
        \put(1,38){\titlestyle{A}}
        \put(66,38){\titlestyle{B}}
    \end{overpic}
    \vspace{-2em}
    \caption{(Color online) Bandstructure of the nano-plasmonic sphere packing illustrated in \figref{spheres}. \titlestyle{A} All $12$ branches, corresponding to the solutions of \eqref{dipoles}. \titlestyle{B} The isolated triplet states that meet at $K_1=0.9992$ and $K_2=1.0076$ show a particularly clear and isotropic Weyl hypercone (red) and a flat dark mode (black) in between, even for relatively large $k \approx \pi/(5a)$.}
    \label{fig:bandstruct}
\end{figure*}

First order degenerate perturbation theory and representation theory, the latter of which provides the selection rules for the matrix elements within the former, allows to derive the slopes of the bandstructure at determinstic points of degeneracy (see supplementary material for details). For deterministic three-fold degeneracies at the $\Gamma$ point, this procedure yields a perturbation matrix $W(\vec{k})$ which is valid for small $k\ll 2\pi/a$ (with lattice constant $a$) and time reversal invariance
\begin{equation}
    W_{\alpha\beta} (\vec{k}) = \imath d \sum_\gamma \epsilon_{\alpha\beta\gamma}\, k_\gamma\text{ ,}
    \label{eq:M_pert}
\end{equation}
with a free parameter $d\in\mathbb{R}$ and $\alpha, \beta, \gamma$ iterating over the three partners of the irreducible representation with $\vec{k}=0$ that span the degenerate eigenspace.
Note the similarity of $W_{\alpha\beta}$ to the Weyl Hamiltonian $\mathcal{W}_{ij}$: the Pauli matrices $\sigma_{ij}^{(\gamma)}$ ($i,j\in\left\{ 1,2 \right\}$) that occur in the latter are here replaced by the 3D Levi-Civita tensor $\epsilon_{\alpha\beta\gamma}$. The first order perturbation eigenvalues corresponding to $W(\vec{k})$ are $k_0^{(1)}:=\omega^{(1)}/c=\{0,\pm d k\}$: they only depend on the absolute value of $\vec{k}$ and describe isotropic hyperconic dispersion.

In the following, we shall define a PWP as the exceptional point $(0,k_0^{(0)})$ at which the two Weyl hypercones $(\vec{k},k_0^{(0)}\pm d k)$ in the four dimensional $(\vec{k},k_0)$ parameter space meet. Although the bandstructure does not support a frequency with vanishing density of states due to the flat band, this definition is justified from a topological perspective: the correlated Chern numbers can be analytically calculated when integrating the Berry curvature over a small sphere in $\vec{k}$ space for each of the three bands. They evaluate to $C=0$ for the flat band and $C=\pm2$ for the two hyperconic bands, showing a non-trivial topological signature, similar to a genuine Weyl point with Chern numbers $C=\pm 1$.

Analysing all 3D space groups \cite{0792365909}, it is straightforward to show that deterministic PWPs at the center of the BZ require chiral cubic symmetry. Interestingly, the trigonal groups $P312$ (149) and $P321$ (150) have two-dimensional representations which split into an anisotropic hypercone if time inversion is present, albeit not at the $\Gamma$ point \cite{2015arXiv151204681C}. A closely related matter is the non-existence of deterministic Dirac points at the $\Gamma$ point of two-dimensional crystals, including the famous honeycomb lattice \cite{PhysRevB.89.134302}.

To demonstrate the predicted behaviour we construct a chiral cubic sphere packing which is minimalistic in the sense that it generates the lowest dimensional vector space possible in models based on for example tight binding or pair interaction. A periodic sphere packing can be constructed by placing spheres on the Wyckoff point of multiplicity $N$ within a given space group \cite{0792365909}. Following this procedure for any non-symmorphic chiral cubic space group $\mathcal{G}$ and Wyckoff multiplicities smaller than $12$ yields a sphere packing that has the symmetry of an achiral supergroup $\mathcal{G}_S$. This counter-intuitive behaviour is related to the isotropy of the sphere as seed object that allows to introduce additional spurious (irregular) rotations.
These unwanted rotations can be suppressed by the finite translation part in non-symmorphic symmetries. A seed sphere on the $4a$ Wyckoff point $(x,x,x)$ of $\mathcal{G}=P2_13$ thus induces a chiral cubic sphere packing with $\mathcal{G}_S=\mathcal{G}$ and only $4$ spheres in the unit cell. Only if $x = n/4$ ($n\in\mathbb{Z}$), the sphere packing acquires the symmetry of the achiral supergroup $Pa\bar{3}$ (205); the chiral supergroup $P4_332$ (212) is induced for $x=1/8+n/2$, and $P4_132$ (213) for $x=-1/8+n/2$. Note that the introduction of an octahedral isogonal point symmetry instead of the tetrahedral symmetry of $P2_13$ in these cases does not impose a change of bandstructure behaviour close to $\Gamma$. \figref{spheres} illustrates the sphere packing for $x=3/8\,a$ ($P4_132$ symmetry).

To elucidate the physics (ahead of a concrete experimental realization) let us consider an effective plasmonic model consisting of metallic nano-spheres of radius $\rho$ in vacuum (as in \cite{Han2009,PhysRevLett.110.106801}). The position $\vec{r}_i$ of sphere $i$ shall be such that the distance $d_{ij}=|\vec{r}_i-\vec{r}_j|\gg \rho$ for any pair of spheres $(i,j)$. In the quasistatic approximation, 
Maxwell's equations take the self-consistent form (acting on the dipole moments ${\vec{p}_i}$) \cite{Draine:94}:
\begin{equation}
    \vec{p}_i = \alpha(k_0)\sum_{j\ne i} \mathcal{G}(\vec{r}_i-\vec{r}_j,k_0)\vec{p}_j\text{ .}
    \label{eq:dipoles}
\end{equation}
Here, $\alpha(k_0)=\rho^3(1-3k_0^2/k_p^2)^{-1}$ is the polarizability of a metallic sphere in vacuum, that is modelled by a non-dissipative Drude response with plasma wave number $k_p$; $\mathcal{G}(\vec{r},k_0)$ is the dyadic Green function for the monochromatic Maxwell wave operator.

If the spheres are arranged periodically as introduced above (\cf \figref{spheres}), the index $i$ is conveniently substituted by a multi-index $(\vec{n},\mu)\in\mathbb{Z}^3\times\{1,2,3,4\}$, with $\vec{r}_{\vec{n},\mu}=\vec{T}_{\vec{n}}+\vec{r}_\mu$ given by the sum of the lattice vector $\vec{T}_{\vec{n}}=a\,\vec{n}$ and the position within the unit cell $\vec{r}_\mu$.
Bloch's theorem then implies $\vec{p}_{\vec{n},\mu} = \vec{p}_\mu \exp\{ \imath \vec{k}\cdot\vec{T}_{\vec{n}} \}$, so that \eqref{dipoles} reduces to a family of $12$-dimensional non-linear Hermitian eigenproblems:
\begin{equation}
    \alpha^{-1}(k_0)\, \vec{p}_\mu = \sum_{\nu} M_{\mu\nu}(\vec{k},k_0)\,\vec{p}_\nu\text{ .}
    \label{eq:eigen}
\end{equation}
Numerical challenges related to the convergence of the lattice sum $M_{\mu\nu}(\vec{k},k_0)$ and their solution are solved in the supplementary material. Since the matrix $M$ generally imposes a small perturbation to the single sphere resonance solution $K_0^{(n)} := \sqrt{3}\,k_0^{(n)}/k_p = 1$ (due to $\rho^{-3}\gg 1$ in \eqref{eigen}), the eigenvalue problem is linearized by approximating $M_{\mu\nu}(\vec{k},k_0)\approx M_{\mu\nu}(\vec{k},k_p/\sqrt{3})=:M_{\mu\nu}(\vec{k})$. This assumption is inadequate close to the Ewald sphere $k_0=|\vec{k}|$, caused by poles in the diagonal entries of $M(\vec{k}=k_0\vec{\hat{k}})$, however, only affecting the two modes at the top and the bottom of the bandstructure on either side of the pole, \cf dashed red line in \figref{bandstruct}\titlestyle{A}.
The eigenvalues $\lambda_n(\vec{k})=\alpha^{-1}(k_0)$ ($n=1,2,\dots,N$) of $M(\vec{k})$ can be obtained numerically with low computational cost. They produce the respective dispersion relation $K_0^{(n)}(\vec{k}) = [1-\rho^3\lambda_n(\vec{k})]^{1/2}$, as shown in \figref{bandstruct} for parameters $x/a=0.175$, $k_0a/(2\pi)=0.1$ and $\rho/a=0.1$.
\figref{bandstruct}\titlestyle{B} shows an example within our model, where the first order perturbation outweighs higher orders even for relatively large Bloch wave number $k\approx\pi/(5a)$, so that an almost perfect hypercone can be observed. We find that, in contrast to this 3D representation, all first order perturbation matrix elements vanish for two-fold degeneracies at the $\Gamma$ point for any space group, cubic or non-cubic ($K_0=0.995$ in \figref{bandstruct}\titlestyle{A} constitutes an example). These exceptional points are henceforth not lifted in first order and are topologically trivial, with Chern number $C=0$ in both bands.

\begin{figure*}[t]
    \hspace{-1em}
    \begin{minipage}[T]{0.67\textwidth}
        \vspace{1em}
        \begin{overpic}[width=\textwidth]{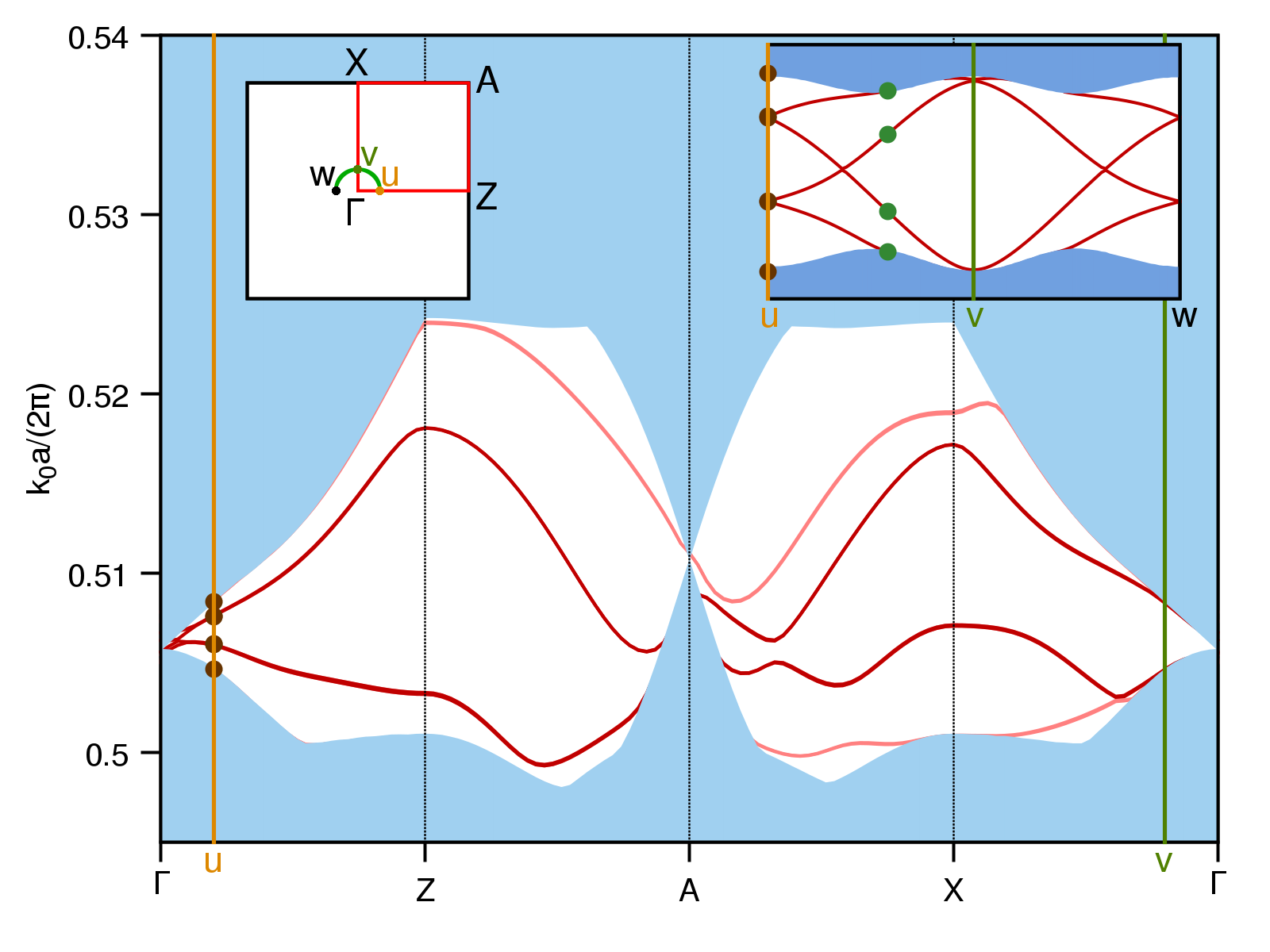}
            \put(1,72){\titlestyle{A}}
            \put(103,72){\titlestyle{B}}
            \put(58,53){\color{Brown}\bf 1}
            \put(58,58){\color{Brown}\bf 2}
            \put(58,65){\color{Brown}\bf 3}
            \put(58,69){\color{Brown}\bf 4}
            \put(67.5,53.3){\color{OliveGreen}\bf 1}
            \put(67.5,57.5){\color{OliveGreen}\bf 2}
            \put(67.5,63.5){\color{OliveGreen}\bf 3}
            \put(67.5,68){\color{OliveGreen}\bf 4}
        \end{overpic}
    \end{minipage}
    \begin{minipage}[T]{0.32\textwidth}
        \vspace{-1em}
        \hspace{10em}
        \rotatebox{90}{
        \begin{overpic}[width=25em]{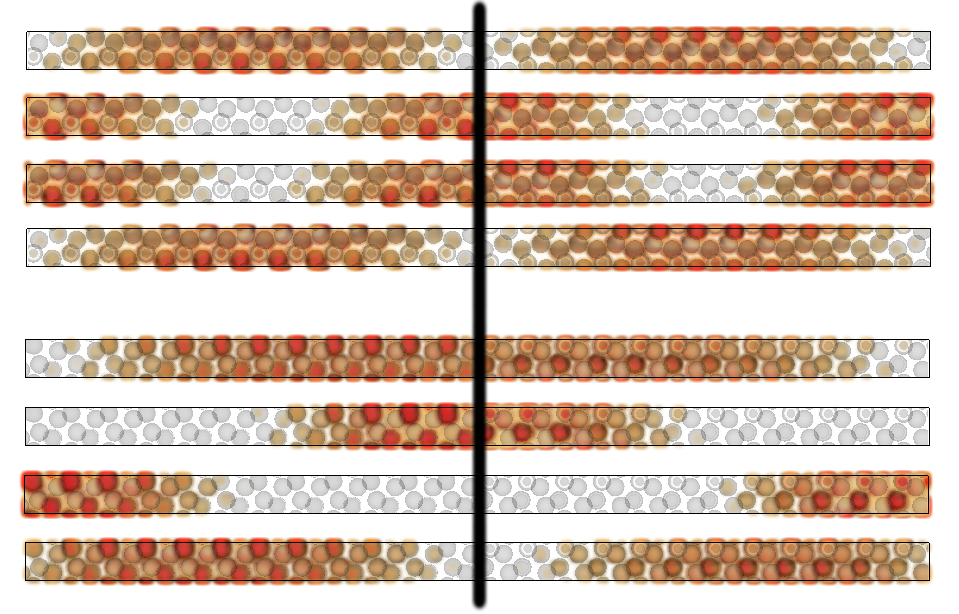}
            \put(99,60){\rotatebox{-90}{\color{Brown}\bf 1}}
            \put(99,53){\rotatebox{-90}{\color{Brown}\bf 2}}
            \put(99,46){\rotatebox{-90}{\color{Brown}\bf 3}}
            \put(99,40){\rotatebox{-90}{\color{Brown}\bf 4}}
            \put(99,28){\rotatebox{-90}{\color{OliveGreen}\bf 1}}
            \put(99,21){\rotatebox{-90}{\color{OliveGreen}\bf 2}}
            \put(99,14){\rotatebox{-90}{\color{OliveGreen}\bf 3}}
            \put(99,7){\rotatebox{-90}{\color{OliveGreen}\bf 4}}
            \put(-1.5,54){\rotatebox{-90}{\scalebox{1.3}{$\rightarrow $}}}
            \put(-2.5,58){\rotatebox{-90}{\scalebox{1.3}{$y$}}}
            \put(12,62){\scalebox{1.4}{$\rightarrow$}}
            \put(8,65.5){\rotatebox{-90}{\scalebox{1.4}{$z$}}}
            \put(0,70){\rotatebox{-90}{\scalebox{1.3}{$x\odot$}}}
            \put(52,75){\rotatebox{-90}{\includegraphics[width=3em]{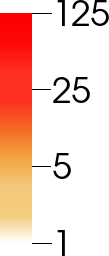}}}
            \put(20,31){\bf RH}
            \put(70,31){\bf LH}
        \end{overpic}}
        \hspace{-3em}
    \end{minipage}
    \vspace{-2em}
    \caption{(Color online) Surface modes close to a PWP frequency. \titlestyle{A} Surface bandstructure for a supercell of two enantiomorphic sphere packings. Topologically protected (dark red) and unprotected (pale red) surface bands are present within the partial gap of the projected bulk bandstructure (blue). The main graph shows the bandstructure along the irreducible BZ boundary (red path, left inset, \cf Fig.~16 in \cite{Setyawan2010299}), whereas the inset on the right follows a small semicircle at the $\Gamma$ point (green path, left inset). The individual paths intersect at two points u and v. \titlestyle{B} Field energy distribution (arbitrary units) for encircled modes of same color in \titlestyle{a}. The interfaces between the right handed (RH) and the left handed (LH) crystal are at the center (black line) and at the end of the unit cell.}
    \label{fig:surface_modes}
\end{figure*}
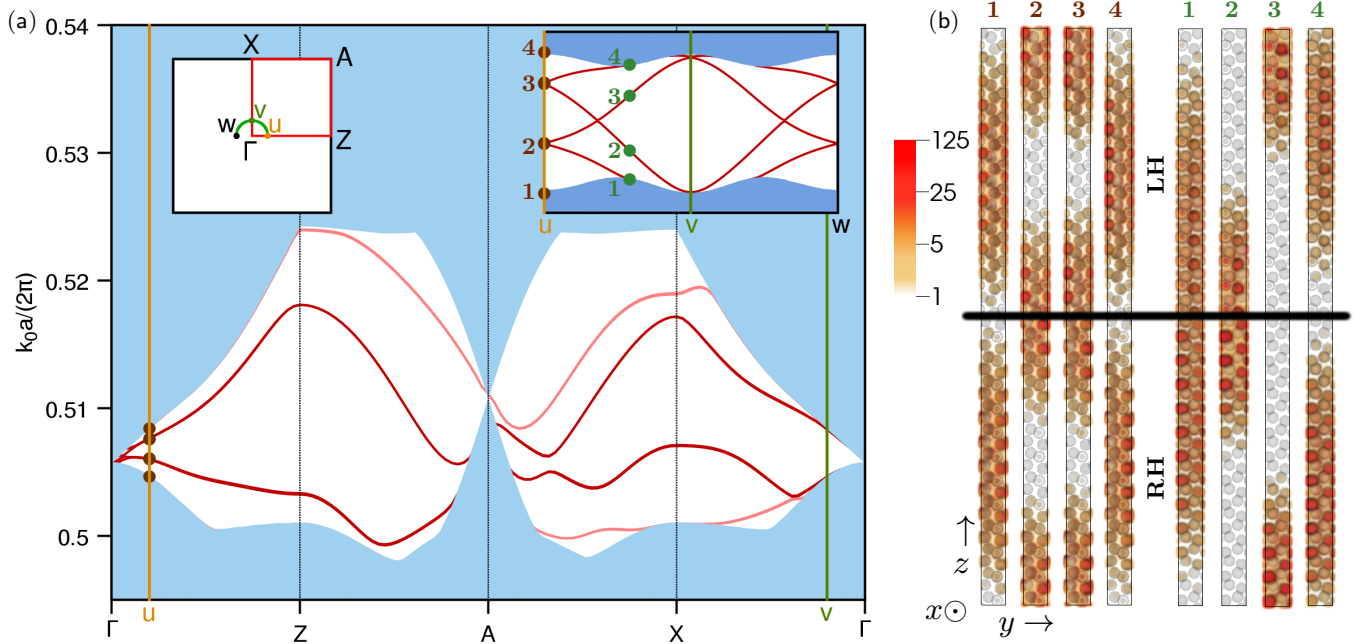

The universality of our results is vividly demonstrated if we replace the small metallic spheres by larger spheres of radius $\rho/a=0.25$ (fill fraction of $\pi/12\approx 26\%$), made of a high refractive index material with $n=4$, thus constructing a photonic crystal analog. The associated bandstructure (calculated with MPB \cite{Johnson2001:mpb}) close to $k_0a/(2\pi)=0.5$ (supplementary figure $1$) resembles \figref{bandstruct}. A partial band gap opens in the projected bulk bandstructure with respect to a $[001]$ inclination in \figref{surface_modes}\titlestyle{A}: this is the blue area of all $(\vec{k}_\parallel,k_0)$ for which at least one bulk mode exists for arbitrary $k_z\in\mathbb{R}$ \cite{JoannopoulosJohnsonWinnMeade:2008}. Since the PWP degeneracy as well as the four-fold degeneracy at $R$ (projected onto $A$) is protected by cubic symmetry, this gap can be opened completely by e.g.~perturbing the sphere positions (supplementary figure $2$).
Topological surface states exist in the band gap at the interface between two enantiomorphic structures (with same bulk bands, but opposite chirality and Chern characteristics): \figref{surface_modes}\titlestyle{A} shows the surface mode dispersion of $12$ unit cells of a right handed crystal ($x/a=0.175$) and $12$ unit cells of a left handed crystal ($x/a=-0.175$) stacked in $[001]$ direction in a supercell geometry. The space group of the supercell is monoclinic with $P2_1/c$ (14) symmetry (note, however, that the Bravais lattice is tetragonal).

The supercell symmetry including time reversal requires all modes along $Z-A-X$ (\figref{surface_modes}\titlestyle{a}, left inset) to be two-fold degenerate. However, the space group representations along $\Gamma-Z$ and $X-\Gamma$ (including $\Gamma$ itself) are 1D. The surface states still stick together. This can be understood as follows: consider a single interface surface mode along $\Gamma-Z$. Along this path, $\vec{k}$ is invariant under the two-fold (screw) rotation $(2)$ in $P2_1/c$ \cite{0792365909} (note that our $x$ axis corresponds to their $y$ axis). This screw rotation further maps the field profile from one interface to the other, so that a 1D representation has to have intensities of equal magnitude on both interfaces. Since the two interfaces are well separated by a zero field bulk region, Maxwell's equations are also satisfied for the same frequency by a field that is non-zero at one of the two interfaces only. The mode must thus be two-fold degenerate. Close to the $\Gamma$ point, the decay length becomes larger than $6$ unit cells, so that the argument breaks down and the degeneracy is lifted. For $X-\Gamma$, the same line of thought applies to the glide plane $(4)$.

\figref{surface_modes}\titlestyle{B} demonstrates that the modes within the bulk gap are indeed localized at the surface, in contrast to the bulk modes within the blue region. There are two degenerate mode pairs at $\vec{k}=0.2\pi/a\times(1,0,0)^T$ (brown points $2$ and $3$ in \figref{surface_modes}\titlestyle{A}, only one states is shown, respectively). The degeneracy splits for $\vec{k}=0.2\pi/a\times(\cos(\phi),\sin(\phi),0)^T$ with arbitrary $\phi$ (green points for $\phi=0.28\pi$).

We have thus shown that surface states exist. But are these also of topological nature? The conventional path $\Gamma{-}Z{-}A{-}X{-}\Gamma$ does not reveal the topological nature of the four surface states emanating from the PWP. To show that these are, indeed, protected, we follow \cite{Wan2011} and consider the cylinder $\vec{k}(\varphi,k_z)=(k\cos(\varphi),k\sin(\varphi),k_z)^T$ (with constant $k$ and $-\pi/a<k_z\le\pi/a$, $0<\varphi\le2\pi$). This cylinder constitutes a closed surface (a torus) in $\vec{k}$-space within which the bandstructure exhibits a band gap, so that a gap Chern number (sum over all bands below the gap) is well defined. The change in gap Chern number $|\Delta C|$ across an interface equals the number of topologically protected surface states that connect the bulk bands below the gap with those above \cite{PhysRevA.78.033834,Lu2014}. The gap Chern number for the above torus and a hyperconic band at a PWP is given by $|C|=2$, as shown above (note that only the Chern number of the PWP at the gap frequency needs to be considered). This results in $8$ surface bands for the supercell geometry with two $|\Delta C|=4$ interfaces, four of which are observed along the half cylinder in \figref{surface_modes}\titlestyle{a}: each of these bands touches and connects the projected bulk bands above and below the gap and thus is, veritably, protected.

In this letter, we have shown that isotropic hyperconic dispersion can be found at the $\Gamma$-point of chiral cubic photonic lattices. The associated exceptional PWP points exhibit the topological characteristics of a double Weyl point. A natural application for the unique dispersion behaviour of these PWPs are zero refractive index materials that have been suggested previously in the context of accidental Dirac points in two-dimensional photonic crystals \cite{Huang2011}. The localization of associated protected surface states and their flatness yield a gigantic density of states, making PWP systems an ideal starting point to explore topological lasing applications \cite{Harari:16} in three dimensions.

\begin{acknowledgements}
    This work was supported by the EPSRC through program grant EP/L027151/1. We would like to thank Dr.~Paloma Arroyo Huidobro, Gleb Siroki and Prof.~Sir John Pendry for helpful discussions.
\end{acknowledgements}

\newpage
\onecolumngrid
    \textbf{\Large
        \begin{center}
            Supplementary Material
        \end{center}
    }
\vspace{1em}
\twocolumngrid
\section{The canonical representation of the plasmonic sphere packing at $\Gamma$ \label{app:canonical_rep}}

The existence of three-fold degeneracies at the $\Gamma$ point in chiral tetrahedral and octahedral space groups is enumerated in the literature \cite{Bradley_GT}. This general results stems from the fact that space group representations at $\Gamma$ can be written as a direct product between the representation of the isogonal point group and the trivial representation of the translation group, \ie
\begin{equation*}
    \Delta_{\vec{k}=0,i} (\left\{ p_\alpha|t_\alpha+T \right\}) := D_i ( p_\alpha )\text{ .}
\end{equation*}
It is straightforward to verify that $\Delta_{\vec{k}=0,i}$ is an irreducible representation of the space group if $D_i$ is an irreducible representation of the isogonal point group. There are therefore 3D irreducible representations ($T$) for any chiral cubic space group associated with $\vec{k}=0$ (rigorous analysis \cite{Bradley_GT} shows that there are no others apart from the above): we tabulate the characters for tetrahedral point group in table \ref{tab:23_characters}, and refer to \cite{PhysRevB.88.245116} for the octahedral group. This generic finding is very useful for the application of the perturbation theory below. Additionally, we here derive the canonical representation of the plasmonic sphere packing and thereby the exact split of the underlying $12$ dimensional vector space into irreducible representations.

\setlength{\extrarowheight}{.1em}
\begin{table}[t]
    \centering
    \begin{ruledtabular}
    \begin{tabular}{l|cccc|c}
        $23$  & $\identity$ & $3C_2$ & $4C_3$     & $4C_3'$    & FS \\\hline
        $A$   & 1           & 1      & 1          & 1          & (a)\\
        $E_+$ & 1           & 1      & $\omega$   & $\omega^2$ & (b)\\
        $E_-$ & 1           & 1      & $\omega^2$ & $\omega$   & (b)\\
        $T$   & 3           & -1     & 0          & 0          & (a)\\
    \end{tabular}
    \end{ruledtabular}
    \caption{Character table for the $23$ tetrahedral point group, with $\omega:=\exp\left\{ 2\imath\pi/3 \right\}$. The last column lists the Frobenius Schur type of the representation.}\vspace{-1em}
    \label{tab:23_characters}
\end{table}

The representation matrix of the vector space $V$ corresponding to different spheres (within the equivalence class of translations) is given by permutation matrices. The following table lists the spheres $\nu$ onto which a sphere $\mu$ is mapped under the symmetry operation $\alpha$ in $P2_13$. The symmetry indices are adapted from \cite{0792365909}.\vspace{1em}

\setlength{\extrarowheight}{.1em}
\begin{ruledtabular}
    \begin{tabular}{|c||c|ccc|cccc|cccc|}
        Class                   &   $E$     &   \multicolumn{3}{c|}{$3C_2$} &   \multicolumn{4}{c|}{$4C_3$} &  \multicolumn{4}{c|}{$4C_3'$}       \\
        $\mu \setminus \alpha$  &   1       &   2   &   3   &   4           &   5   &   6   &   7   &   8   &   9   &   10  &   11  &   12      \\
        \hline
        1                       &   1       &   2   &   3   &   4           &   1   &   4   &   2   &   3   &   1   &   3   &   4   &   2       \\   
        2                       &   2       &   1   &   4   &   3           &   4   &   1   &   3   &   2   &   3   &   1   &   2   &   4       \\   
        3                       &   3       &   4   &   1   &   2           &   2   &   3   &   1   &   4   &   4   &   2   &   1   &   3       \\   
        4                       &   4       &   3   &   2   &   1           &   3   &   2   &   4   &   1   &   2   &   4   &   3   &   1       \\
        \hline
        Trace $\chi$            &   4       &   \multicolumn{3}{c|}{0}      &   \multicolumn{4}{c|}{1}      &   \multicolumn{4}{c|}{1}
    \end{tabular}
\end{ruledtabular}\vspace{1.5em}

The space within which the polarisation vector is rotated is the three-dimensional Euclidean space $W$. We only list the relevant trace (or character) of the rotation matrix $\chi(\phi)=1+2\cos(\phi)$ here, where $\phi$ denotes the angle of rotation. The characters of the symmetry operations in $V$ and $W$, and of the canonical representation for the system vector space $V\otimes W$ are summarised below.\vspace{1.5em}

\setlength{\extrarowheight}{.1em}
\begin{ruledtabular}
    \begin{tabular}{|c||c|c|c|c|}
        Class               &   $E$     &   $3C_2$ &   $4C_3$ &  $4C_3'$       \\
        \hline
        $\chi^V$            &   4       &   0       &   1      &   1            \\
        $\chi^W$            &   3       &   -1      &   0      &   0            \\
        \hline
        $\chi^{V\otimes W}$ &   12      &   0       &   0      &   0
    \end{tabular}
\end{ruledtabular}\vspace{1.5em}

The characters of the irreducible representations of the associated tetrahedral point group are shown in \tabref{23_characters}. Note that we use the Schoenflies notation $23$ in this section in order to avoid confusion with the symbol $T$ that is conventionally used for three-dimensional irreducible representations. The canonical representation at the $\Gamma$ point reduces into
$$A+E_++E_-+3T$$
(where $(+)$ denotes a direct sum operation in this context) according to the reduction formula \cite{Dresselhaus}
\begin{equation}
    n_i = \frac{1}{N} \sum_R \cc{\chi}(R)\,\chi_i(R)\text{ ,}
    \label{eq:reduction}
\end{equation}
where $n_i$ quantifies how often an irreducible representation $i$ is contained in the reducible representation with characters $\chi(R)$ ($\cc{\cdot}$ denotes complex conjugation), $N$ is the number of elements $R$ in the group, and $\chi_i(R)$ is the character of the irreducible representation.

Note that the representation $E_\pm$ form a two-fold degenerate pair $E$ because the system is also invariant under time inversion, as evident from the Schur type $(b)$ in \tabref{23_characters} here. For details on time inversion symmetry in addition to space group symmetries see for example Herring \cite{PhysRev.52.361} and Frobenius \cite{Frobenius}, page 354 ff. The predicted split is thus reproduced in Fig.~2 in the main manuscript: the bandstructure shows one trivial $A$ representation, one 2D $E$ representation, and three 3D $T$ representations, two of which produce genuine hyperconic dispersion, while the third is degenerate in first order and corresponds to the special case $d=0$ below.

\section{Perturbation theory and selection rules \label{app:pert}}

Reduction into the irreducible representations of the respective subgroup of the space group $\mathcal{G}$ as described in \cite{PhysRevB.88.245116} predicts how and whether a degenerate mode splits when going away from that point. If not perturbed along a high symmetry direction in reciprocal space it always splits into 1D representations as the Bloch representations, characterized by $\vec{k}$, with respect to the invariant subgroup of translations generate a full star, with the same dimension as the underlying point group.

We are here however concerned with the order of splitting: for linear Weyl dispersion, the first order cannot vanish. We here derive an approach that is analogous to, but more general than the well established $\vec{k}\cdot\vec{p}$ perturbation theory \cite{Dresselhaus}. Consider a generic non-linear eigenproblem
\begin{equation}
    M(\lambda) \vec{v} = 0\text{ ,}
    \label{eq:generic_eigen}
\end{equation}
characterized by an operator $M(\lambda)$ that commutes with all operators of $\mathcal{G}$. The eigenvectors $\vec{v}$ are partners $(\vec{k}_n,\alpha)$ of the irreducible representations of the space group enumerated by $(\vec{k},i)$. In this context, $\vec{k}$ is a Bloch wave vector in the irreducible Brillouin zone and $i$ denotes different induced representations from the small representations of the associated little group, whereas $\vec{k}_n$ iterates over different representatives in the star of $\vec{k}$ and $\alpha$ over partners of the small representations of the little group. We hence write:
\begin{equation*}
    \vec{v} = \vec{v}^{(i)}_{\alpha}(\vec{k},\vec{k}_n)\text{ .}
\end{equation*}
Working in the spatial Fourier domain, we can suppress the explicit $\vec{k}_n$ dependence in the eigenvector, dividing by the phase factor $\exp\left\{ \imath \vec{k}_n\cdot\vec{r} \right\}$ corresponding to a translation  group representation. At an arbitrary point within the BZ we define
\begin{equation*}
    \exp\left\{ -\imath \vec{k}_n\cdot\vec{r} \right\} \vec{v}^{(i)}_{\alpha}(\vec{k},\vec{k}_n) \equiv |\vec{k}i,n\alpha\rangle\text{ .}
\end{equation*}
Importantly, these are partners $\alpha$ of the irreducible representations of the little group at $\vec{k}_n$. The eigenvalue equation becomes
\begin{equation*}
    M(\vec{k}_n,\lambda) |\vec{k}i,n \alpha\rangle = 0\text{ .}
\end{equation*}
For $\vec{k}_n^{(1)} = \vec{k}_n^{(0)}+\delta\vec{k}$, with small $|\delta\vec{k}| \ll \pi/a$, and hence $\lambda_1=\lambda_0+\delta\lambda$, a first order Taylor expansion of $M$ around $M(\vec{k}_n^{(0)},\lambda_0)$ and an expansion of the eigenfunctions into the aforementioned partners $\alpha$ (\cf representation theorem (iii) in \cite{PhysRevB.88.245116}) yields:
\begin{align}
    \Big( M(\vec{k}_n^{(0)},\lambda_0) & \notag \\
    + \delta\vec{k}\cdot\left[\nabla_{\vec{k}}M(\vec{k},\lambda_0)\right]_{\vec{k}=\vec{k}_n^{(0)}} &  \notag\\
    + \delta\lambda\left[\partial_{\lambda}M(\vec{k}_n^{(0)},\lambda)\right]_{\lambda=\lambda_0} \Big) &  \notag\\
        \sum_\alpha c_\alpha |\vec{k}i,n\alpha \rangle & = 0\text{ ,}
        \label{eq:pert}
\end{align}
where we have assumed that $i$ labels the irreducible representation corresponding to the eigenvalue $\lambda_0$ and used the fact that $M(\vec{k}_n^{(0)},\lambda_0)|\vec{k}j,n\beta\rangle\ne 0$ for any representation $j\ne i$ (\ie~we exclude accidental degeneracies), so that a finite coefficient would contradict \eqref{pert} for arbitrary but small $\delta\vec{k}$. In the weak form, \eqref{pert} reads:
\begin{equation}
    \sum_\beta\vec{W}_{\alpha\beta}\cdot\delta\vec{k}\,c_\beta = -\sum_\beta \delta\lambda\, E_{\alpha\beta} c_\beta\text{ ,}
    \label{eq:pert_weak}
\end{equation}
with
\begin{equation*}
    \vec{W}_{\alpha\beta} := \langle \vec{k}i,n\alpha| \left[\nabla_{\vec{k}}M(\vec{k},\lambda_0)\right]_{\vec{k}=\vec{k}_n^{(0)}}|\vec{k}i,n\beta\rangle
\end{equation*}
and
\begin{equation*}
    E_{\alpha\beta} := \langle \vec{k}i,n\alpha| \left[\partial_{\lambda}M(\vec{k}_n^{(0)},\lambda)\right]_{\lambda=\lambda_0} |\vec{k}i,n\beta\rangle\text{ .}
\end{equation*}
The partial derivative $\partial\lambda$ is trivially invariant under all operations $P$ of the little group at $\vec{k}_n^{(0)}$. $M$ is invariant under all space group operations, and $P\in\mathcal{G}$, while $P|\vec{k}i,n\alpha\rangle = \sum_\beta D_{\alpha\beta}(P)|\vec{k}i,n\beta\rangle$. Hence, the action on $W$ and $E$ is generally given by
\begin{align*}
    P E_{\alpha\beta} &= \sum_{\gamma\delta} \cc{D}_{\alpha\gamma}(P) D_{\beta\delta}(P)\, E_{\gamma\delta} \\
    P W^{(n)}_{\alpha\beta} &= \sum_{\gamma\delta,m} \cc{D}_{\alpha\gamma}(P) D_{\beta\delta}(P) R_{nm}(P)\, W^{(m)}_{\gamma\delta}\text{ ,}
\end{align*}
where $R(P)$ denotes the Rodrigues' matrix representation that acts in the standard 3D Euclidean vector space. Since $\delta\vec{k}$ is arbitrary, according to \eqref{pert_weak}, $\vec{W}$ and $E$ are also required to be invariant under all $P$. Introducing the respective direct product space, $E$ and $W$ must hence for all $P$ be in the kernel of $\Delta(P)$ and $\Lambda(P)$, respectively, where
\begin{align}
    \Delta_{(\alpha\beta),(\gamma\delta)}(P) :=& \cc{D}_{\alpha\gamma}(P) D_{\beta\delta}(P) - \delta_{\alpha\gamma}\delta_{\beta\delta}\label{eq:ker_del}\\
    \Lambda_{(\alpha\beta n),(\gamma\delta m)}(P) := &\cc{D}_{\alpha\gamma}(P) D_{\beta\delta}(P) R_{nm}(P)\notag\\
    &- \delta_{\alpha\gamma}\delta_{\beta\delta}\delta_{nm}\label{eq:ker_del}\text{ .}
\end{align}
This reveals the group theoretically allowed form of $E$ and $W$, substantially reducing their degrees of freedom and providing selection rules for many of their entries. $E$ and $W$ are respectively non-zero only if $\Delta+\identity$ and $\Lambda+\identity$ contain the trivial representation if decomposed into the irreducible representations of $i$ via \eqref{reduction} (representation theorem (ii) in \cite{PhysRevB.88.245116}), \ie
\begin{align}
    \frac{1}{\sum_P}\sum_P \cc{\chi_D}(P)\chi_D(P) &\ne 0 \label{eq:char_del}\\
    \frac{1}{\sum_P}\sum_P \cc{\chi_D}(P)\chi_D(P)\text{Tr}\{R(P)\} &\ne 0 \text{ .}
    \label{eq:char_lam}
\end{align}
More precisely, the integer number on the left side of \eqref{char_del} and \eqref{char_lam} is equal to the dimension of the intersection of the nullspaces of $\Delta$ and $\Lambda$, respectively, and hence specifies the degrees of freedom in $E$ and $W$ (but not their form).

\section{Perturbation theory at $\Gamma$ \label{app:pert_Gamma}}
Consider now a perturbation around the centre of the BZ using the equations derived in the previous section. The isogonal little group is here the isogonal point group of the lattice, \ie for chiral cubic symmetry either tetrahedral ($23$) or octahedral ($432$). We here discuss the tetrahedral symmetry, but note that the main results regarding PWPs are exactly the same for the octahedral point group. Consider first a non-degenerate mode with trivial irreducible representation $A$ (see table \ref{tab:23_characters}). While \eqref{char_del} is satisfied, \eqref{char_lam} is not. $W$ is therefore trivial, with all its elements vanishing and \eqref{pert_weak} yielding $\delta\lambda = 0$. A non-degenerate mode is henceforth flat at the centre of the BZ in a cubic symmetry, even in the absence of inversion or time reversal symmetry. Note, that the $E_\pm$ modes show the same behaviour, so that a deterministic two-fold degenerate mode, as present in the case of time reversal only in tetrahedral symmetries is also predicted to be flat at the centre of the BZ. Further note that the two-fold $E$ representation in the octahedral point group (table 3.33 in \cite{Dresselhaus}) also yields a flat band.

The representation $T$ in table \ref{tab:23_characters} (the Rodrigues' representation itself) that corresponds to a three-fold degeneracy, however, satisfies both \eqref{char_del} and \eqref{char_lam}, with $\text{dim}(\text{ker}(\Delta))=1$ and $\text{dim}(\text{ker}(\Lambda))=2$. We compute the intersection of the kernels of $\Delta(P)$ and $\Lambda(P)$ for all $P$ numerically, and find explicitly for the $T$ degeneracy that $E$ is a scalar matrix
\begin{align*}
    E = c \identity
\end{align*}
and that $\vec{W}$ is given by
\begin{equation*}
    \vec{W} = d_1
    \begin{pmatrix}
        \begin{pmatrix}
            0 & 0 & 0 \\
            0 & 0 & 0 \\
            0 & 1 & 0
        \end{pmatrix} \\
        \begin{pmatrix}
            0 & 0 & 1 \\
            0 & 0 & 0 \\
            0 & 0 & 0
        \end{pmatrix} \\
        \begin{pmatrix}
            0 & 0 & 0 \\
            1 & 0 & 0 \\
            0 & 0 & 0
        \end{pmatrix} \\
    \end{pmatrix} 
    + d_2
    \begin{pmatrix}
        \begin{pmatrix}
            0 & 0 & 0 \\
            0 & 0 & 1 \\
            0 & 0 & 0
        \end{pmatrix} \\
        \begin{pmatrix}
            0 & 0 & 0 \\
            0 & 0 & 0 \\
            1 & 0 & 0
        \end{pmatrix} \\
        \begin{pmatrix}
            0 & 1 & 0 \\
            0 & 0 & 0 \\
            0 & 0 & 0
        \end{pmatrix} \\
    \end{pmatrix}\text{ .}
\end{equation*}
In the presence of time reversal symmetry we further require that $\cc{M}(-\vec{k})=M(\vec{k})$, and hence $c\in\mathbb{R}$ and $W_\gamma$ Hermitian, so that \eqref{pert_weak} corresponds to the perturbation matrix given in eq.~$1$ of the main manuscript:
\begin{equation*}
     \imath d \sum_{\beta\gamma} \varepsilon_{\alpha\beta\gamma}\, \delta k_\gamma\,c_\beta = \delta\lambda\, c_\alpha\text{ ,}
\end{equation*}
where we set \wolog $d_1=-\imath d$, $d_2=\imath d$ with $d\in\mathbb{R}$ and $c=-1$.

\section{Chern numbers for the hyperconic bands \label{app:Chern}}

We here calculate the Chern numbers for the three bands that meet at the PWP. We start from the perturbation matrix (Eq.~$1$ in the main manuscript, derived above)
\begin{equation*}
    W_{\alpha\beta} (\vec{k}) = \imath\,d \sum_\gamma \epsilon_{\alpha\beta\gamma}\, k_\gamma\text{ .}
\end{equation*}
To study the topological nature of the PWP, it is convenient to calculate the Chern number over a small sphere in reciprocal space, with the PWP at its centre. We first introduce spherical coordinates with $\vec{k}:=k(\cos{\phi}\sin{\theta},\sin{\phi}\sin{\theta},\cos{\theta})^T$ and associated basis vectors $\{\vec{\hat{k}},\vec{\hat{\theta}},\vec{\hat{\phi}}\}$. With the eigenvectors expanded as $\vec{v}=v_k\vec{\hat{k}}+v_\theta\vec{\hat{\theta}}+v_\phi\vec{\hat{\phi}}$, the perturbation equation
\begin{equation*}
    \imath d k\, \vec{\hat{k}}\times\vec{v} = \lambda \vec{v}
\end{equation*}
has the solution set
\begin{equation*}
    \left[ \lambda,
        \begin{pmatrix}
            v_k \\ v_\theta \\ v_\phi
        \end{pmatrix}
    \right] = 
    \left\{ 
        \left[ 
            0,
            \begin{pmatrix}
                1 \\ 0 \\ 0
            \end{pmatrix}
        \right],
        \left[ 
            \pm dk, \frac{1}{\sqrt{2}}
            \begin{pmatrix}
                0 \\ 1 \\ \pm\imath
            \end{pmatrix}
        \right]
    \right\}\text{ .}
\end{equation*}
We note that $\vec{v}$ above is the vector over the coefficients $c_\alpha$ of partners $|\alpha\rangle$ of the irreducible $T$ representation. These are by definition orthonormalized and $\vec{k}$ independent. Topologically, the discussion of any physical Hilbert space therefore collapses to the three-dimensional vector space of the coefficients. The Gaussian curvature
\begin{equation*}
    K = dt_1\wedge dt_2\,\left[(\partial_2\vec{e}_1,\partial_1\vec{e}_2) - (\partial_1\vec{e}_1,\partial_2\vec{e}_2)\right]
\end{equation*}
is a measure of intrinsic curvature of the tangent space of a surface spanned by local basis vectors $\vec{e}_1$ and $\vec{e}_2$. The Berry curvature is a similar measure concerning a generic complex vector field $|v(\vec{x})\rangle$ that is attached to a manifold $\mathcal{M}$ parametrized by $\vec{x}$. It can be conveniently written as the two-form
\begin{figure}[t]
    \centering
    \includegraphics[width=.48\textwidth]{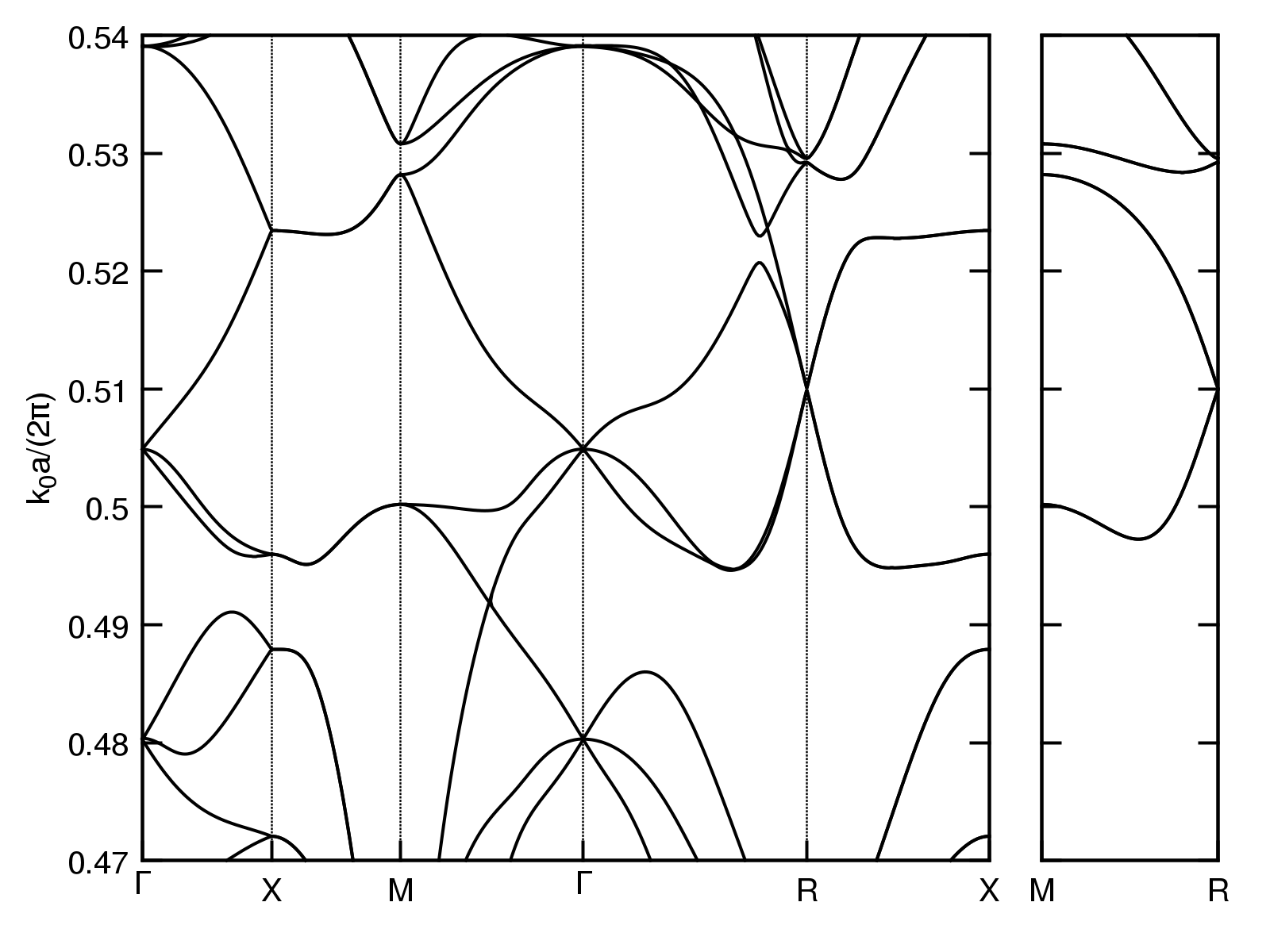}
    \caption{Bulk bandstructure for the photonic crystal sphere packing with same parameters as in figure $3$ of the main manuscript.}
    \label{fig:bulk}
\end{figure}
\begin{equation*}
    \mathcal{B} = \imath \text{Tr}\{P dP\wedge dP\}\text{ ,}
\end{equation*}
with the projector $P:=|v\rangle\langle v|$ and its exterior derivative $dP=\sum_i dx_i \partial_{x_i}P$ in the tangent space of $\mathcal{M}$. Note that the wedge product does not vanish as $P$ is an operator, but the Berry curvature is easily shown to be real. In our case, the manifold is 2D, with $x_1=\theta$ and $x_2=\phi$, so that the Berry curvature reduces to
\begin{equation}
    \mathcal{B}
    = 2\, d\theta \wedge d\phi \, \imag{\langle \partial_\phi v | \partial_\theta v \rangle} \text{ .}
    \label{eq:Berry}
\end{equation}
Similar to the Euler number $\chi:=(2\pi)^{-1}\int_\mathcal{M} K$ that approaches integer values for any closed surface and is directly related to its genus, the Chern characteristics
\begin{equation*}
    C = \frac{1}{2\pi}\oint_\mathcal{M} \mathcal{B}
\end{equation*}
is also quantized and constitutes a topological invariant of the vector field. The flat band eigenvector $\vec{v}=\vec{\hat{k}}$ is real and trivially leads to a vanishing Berry curvature when substituted into \eqref{Berry}. The eigenvectors of the hyperconic bands $\vec{v}_\pm = (\vec{\hat{\theta}} \pm \imath \vec{\hat{\phi}})/\sqrt{2}$ on the other hand yield a Berry curvature of
\begin{equation*}
    \mathcal{B} = \pm\, d\theta \wedge d\phi \, \left( \partial_\theta \vec{\hat{\theta}}, \partial_\phi \vec{\hat{\phi}} \right) = \pm \sin(\theta)\, d\theta \wedge d\phi\text{ ,}
\end{equation*}
and hence a Chern characteristics of $C=\pm 2$, in direct analogy to the Euler number $\chi=2$ of the sphere.

\section{Additional figures}

\begin{figure}[t]
    \centering
    \includegraphics[width=.48\textwidth]{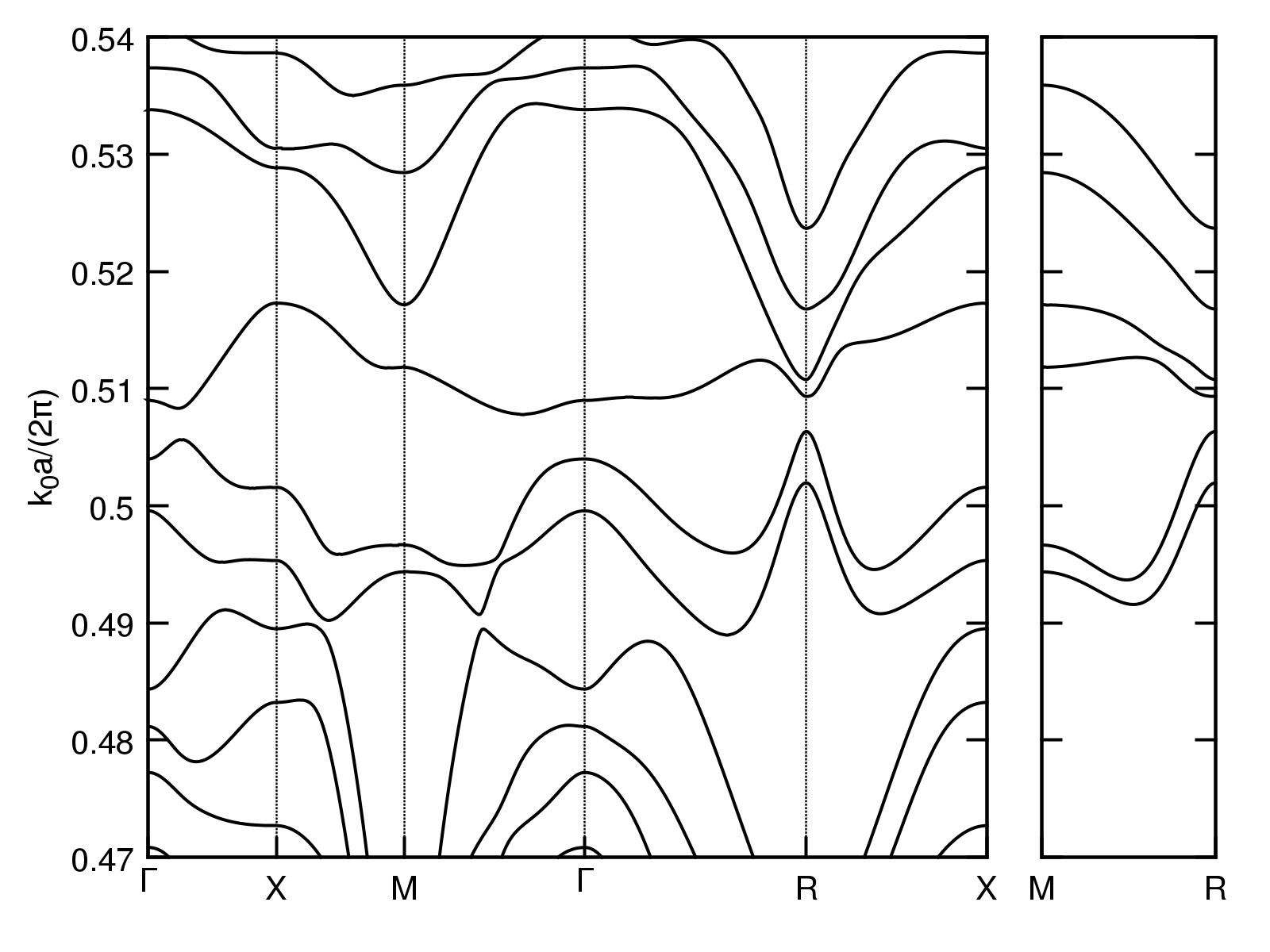}
    \caption{Bulk bandstructure for a perturbed photonic crystal sphere packing with same parameters as in figure $3$ of the main manuscript, except that the sphere positions are shifted by random vectors $\delta\vec{r}$, with $r_i$ uniformly distributed between $0$ and $0.1a$.}
    \label{fig:pert}
\end{figure}

\figref{bulk} shows the plain bulk bandstructure of the photonic crystal sphere packing close to the PWP of interest around $k_0a/(2\pi)=0.5$. In \figref{pert} the sphere positions in the unit cell are distorted, thereby destroying cubic symmetry and both degeneracies at the $\Gamma$ and at the $R$ point. We note that, despite the fact that the Bravais lattice remains simple cubic by construction, the path followed in \figref{pert} is thus in principle inappropriate, but kept for the sake of comparability with \figref{bulk}. A full bandgap is, however, observed for the particular realization in the whole (triclinic) Brillouin zone. Opening the gap does not change the topological characteristics of the bands above and below the gap. Frequency isolated protected surface states can hence be found for the perturbed structure.

\section{The eigenvalue equation for the sphere packing \label{app:spheres}}

Since a proper derivation of Eq.~1 of the main manuscript,  reproduced here for clarity
\begin{equation}
    \vec{p}_i = \alpha(k_0)\sum_{j\ne i} \mathcal{G}(\vec{r}_i-\vec{r}_j,k_0)\vec{p}_j\text{ ,}
    \label{eq:dipoles}
\end{equation}
seems to be absent in the literature, we here rigorously show its validity in the assumed limits.
We model the dielectric matrix intrinsically via an effective displacement field with the corresponding permittivity $\varepsilon_d$, and the polarisation $\vec{P}_m$ of the metallic spheres explicitly. The macroscopic Maxwell equations for monochromatic fields can then be transformed into a wave equation for the electric field $\vec{E}$ and the polarisation field $\vec{P}_m$, that vanishes in the dielectric matrix:
\begin{equation}
    \underbrace{-(4\pi k_0^2)^{-1}\, \left( \kappa^2 - \text{curl}^2 \right)}_{:=\mathcal{M}}\vec{E} = \vec{P}_m\text{ .}
    \label{eq:Maxwell}
\end{equation}
We have here introduced the wave number $\kappa:=\sqrt{\varepsilon_d} k_0$ and the Maxwell operator $\mathcal{M}$ for convenience.

\eqref{Maxwell} can be inverted using the dyadic Green function
\begin{equation}
    \mathcal{G}_\mathcal{M}(\vec{r}) = -4\pi k_0^2\, \left( \identity + \frac{\nabla\otimes\nabla}{\kappa^2} \right) \mathcal{G}_\mathcal{H}(r)
    \label{eq:Green}
\end{equation}
that solves the equation $\mathcal{M}\mathcal{G}_\mathcal{M}(r) = \identity\delta^{(3)}(\vec{r})$, with the (retarded) Helmholtz Green function $\mathcal{G}_\mathcal{H}(r) = -(4\pi r)^{-1}\exp\left\{ \imath\kappa r \right\}$ that solves
$$\left( \kappa^2 + \Delta \right)\mathcal{G}_\mathcal{H} = \delta^{(3)}(\vec{r})\text{ .}$$
We assume that the metallic material response $\varepsilon_m(k_0)$ is known and frequency dependent. Therefore, with $\vec{E} = 4\pi/(\varepsilon_m-\varepsilon_d)\,\vec{P}_m$, \eqref{Maxwell} transforms into a relation on the metallic subdomain only:
\begin{equation}
    \frac{4\pi}{\varepsilon_m-\varepsilon_d}\,\vec{P}_m(\vec{r}) = \int d^3r'\,\mathcal{G}_\mathcal{M}(\vec{r}-\vec{r}')\,\vec{P}_m(\vec{r}')
    \label{eq:inverseMaxwell}
\end{equation}
The polarisation field inside a single sphere can be considered constant for sufficiently small frequencies, in zero order in $k_0\rho \ll 1$.\footnote{Note that this condition does not impose a strong constraint on $k_0$ itself, since we assume small spheres $\rho\ll \min(d_{\mu\nu}) < a$.} We hence make the ansatz
\begin{equation*}
    \vec{P}_m(\vec{r}) = \frac{1}{V_S}\sum_i \vec{p}_i \chi(\vec{r}-\vec{r}_i)\;\text{ , with }\chi(\vec{r})=
    \begin{cases}
        1 & \text{if } |\vec{r}| < \rho\\
        0 & \text{else}
    \end{cases}
\end{equation*}
and $V_S=(4\pi\rho^3)/3$ the sphere volume. Substitution into \eqref{inverseMaxwell} and testing with the basis functions $\chi(\vec{r}-\vec{r}_i)$ yields an equation for the dipole moments $\vec{p}_i$ only:
\begin{equation}
    \frac{3\rho^3}{\varepsilon_m-\varepsilon_d} \vec{p}_i = \sum_j I_\rho(\vec{r}_i,\vec{r}_j)\,\vec{p}_j\text{ .}
    \label{eq:invpolMaxwell}
\end{equation}
Here,
\begin{equation*}
    I_\rho(\vec{r}_i,\vec{r}_j) = \frac{1}{V_S^2}\int_{\mathcal{B}_\rho(\vec{r}_i)}d^3r\;\int_{\mathcal{B}_\rho(\vec{r}_j)}d^3r'\;\mathcal{G}_\mathcal{M}(\vec{r}-\vec{r}')
\end{equation*}
is a double integral over balls $\mathcal{B}_\rho(\vec{r})$ with radius $\rho$ at position $\vec{r}$. For any two spheres $i\ne j$ and in the limit of small spheres $\rho \ll d_{ij}$, $I_\rho(\vec{r}_i,\vec{r}_j)$ reduces to $V_S\, \mathcal{G}_\mathcal{M}(\vec{r}_i-\vec{r}_j)$. For $i=j$, we use the following theorem.

\noindent{\bf Theorem I:} Let
\begin{equation*}
    I = \int_{\mathcal{B}_\rho}d^3r\;\int_{\mathcal{B}_\rho}d^3r'\; f(\vec{x})
\end{equation*}
be a double integral over balls (at the origin) $\mathcal{B}_\rho$ of a function $f(\vec{x})$ that only depends on the difference in coordinates $\vec{x}=\vec{r}-\vec{r}'$. The integral can then be expressed as a single integral over a Ball $\mathcal{B}_{2\rho}$
\begin{equation*}
    I = \frac{\pi}{12}\int_{\mathcal{B}_{2\rho}}d^3x\;(4\rho+x)(2\rho-x)^2\,f(\vec{x})\text{ .}
\end{equation*}\\
\noindent{\bf Proof:} A change of coordinates $(\vec{r},\vec{r}')\mapsto(\vec{x},\vec{R}:=\vec{r}+\vec{r}')$ yields:
\begin{equation*}
    I = \frac{1}{8} \int_{\mathcal{B}_{2\rho}}d^3x\;f(\vec{x})\underbrace{\int_{\mathcal{V}(\vec{x})}d^3R}_{=:V[\mathcal{V}(\vec{x})]}\text{ ,}
\end{equation*}
where, with the binary operator $(+)$ on spatial domains denoting the Minkowski sum,
\begin{equation*}
    \mathcal{V}(\vec{x}) = \left( \mathcal{B}_\rho \cap \mathcal{B}_\rho(\vec{x}) \right) + \left( \mathcal{B}_\rho \cap \mathcal{B}_\rho(-\vec{x}) \right)\text{.}
\end{equation*}
Using the translation invariance of the volume and a mixed volume expansion \cite{burago1988geometric}, we arrive at:
\begin{align*}
    V[\mathcal{V}(\vec{x})] &= V\left[ \left( \mathcal{B}_\rho \cap \mathcal{B}_\rho(\vec{x}) \right) + \left( \mathcal{B}_\rho \cap \mathcal{B}_\rho(-\vec{x}) \right) \right] \\
                   &= V\left[ \left( \mathcal{B}_\rho \cap \mathcal{B}_\rho(\vec{x}) \right) + \left( \mathcal{B}_\rho \cap \mathcal{B}_\rho(\vec{x}) \right) \right] \\
                   &= 8 V\left[ \mathcal{B}_\rho \cap \mathcal{B}_\rho(\vec{x}) \right] \\
                   &= \frac{2\pi}{3}\left( 2\rho-x \right)^2 \left( 4\rho+x \right) \text{ .}
\end{align*}
In the last line we used that the volume is twice that of a spherical cap with sphere radius $\rho$ and height $\rho-x/2$. The proposition follows from resubstitution. $\square$

Applying theorem I with $f(\vec{x}) = \mathcal{G}_\mathcal{M}(\vec{x})$ and the low frequency condition on the Green function yields the simplified expression
\begin{equation*}
    \mathcal{G}_\mathcal{M}^{(\alpha\beta)}(\vec{x}) = \varepsilon_d^{-1}\partial_\alpha\partial_\beta\frac{1}{x}\text{ ,}
\end{equation*}
with only two independent entries $\mathcal{G}_\mathcal{M}^{(zz)}$ and $\mathcal{G}_\mathcal{M}^{(xz)}$. The integral over the latter entry vanishes as the integrand is antisymmetric in both the azimuth $\phi$ and the polar angle $\theta$. The former integral evaluates to the simple expression
\begin{equation*}
    I^{(zz)}(\vec{r}_i,\vec{r}_i) = -\rho^{-3}\,\varepsilon_d^{-1}\text{ .}
\end{equation*}
Substitution of these results into \eqref{invpolMaxwell} yields
\begin{equation}
    \underbrace{\frac{\varepsilon_m+2\varepsilon_d}{\varepsilon_m-\varepsilon_d}\rho^{-3}}_{:=\alpha^{-1}}\, \vec{p}_i = \varepsilon_d \sum_{j\ne i} \mathcal{G}_\mathcal{M} (\vec{r}_i-\vec{r}_j)\vec{p}_j\text{ .}
\end{equation}
For spheres made of a lossless Drude material $\varepsilon_m(k_0)=1-k_p^2/k_0^2$, embedded in vacuum $\varepsilon_d=1$, this equation is identical to \eqref{dipoles}. We note that this describes an assembly of point dipoles with polarisability $\alpha$, that are interacting via the Maxwell Greens function without self-interaction. This simple picture apparently describes the correct physics in the assumed limits, although it is not possible to derive \eqref{dipoles} within its scope.

\section{The lattice sum}

Substituting the Bloch form $\vec{p}_{\vec{n},\mu} = \vec{p}_\mu \exp\left\{ \imath\vec{k}\cdot\vec{T}_{\vec{n}} \right\}$ into \eqref{dipoles} yields
\begin{equation*}
    \sum_{m,\nu}\mathcal{G}(\vec{r}_\mu-\vec{r}_\nu+\vec{T}_{\vec{n}}-\vec{T}_m)\,e^{\imath\vec{k}\cdot(\vec{T}_m-\vec{T}_{\vec{n}})}\,\vec{p}_\nu = \alpha^{-1}\,\vec{p}_\mu\text{ ,}
\end{equation*}
where we define $\mathcal{G}(0):=0$ to simplify the notation.

Eq.~2 from the main manuscript, namely
\begin{equation}
    \alpha^{-1}(k_0)\, \vec{p}_\mu = \sum_{\nu} M_{\mu\nu}(\vec{k},k_0)\,\vec{p}_\nu\text{ ,}
    \label{eq:eigen}
\end{equation}
results from a lattice vector shift $\vec{T}_m\mapsto\vec{T}_m+\vec{T}_{\vec{n}}$, with the matrix
\begin{equation}
    M_{\mu\nu}(\vec{k}) := \sum_{\vec{n}} \mathcal{G}_\mathcal{M}(\vec{r}_\mu-\vec{r}_\nu-\vec{T}_{\vec{n}})e^{\imath\vec{k}\cdot\vec{T}_{\vec{n}}}\text{ .}
    \label{eq:eigen_matrix}
\end{equation}

This numerical lattice sum that is truncated by for example $|\vec{n}|<N$ converges very slowly to the infinite series \eqref{eigen_matrix}.\footnote{Precisely speaking, it does not converge at all in the vicinity of Ewald's sphere $|\vec{k}|=k_0$ due to a ringing like phenomenon (spatial and temporal dispersion counteract and prevent the summands from alternating in one direction).} Convergence is generally guaranteed by the alternating phase factor, but the rate of convergence is governed by the long range $r^{-3}$ interaction from the Green function, and can only be algebraic. This problem is well known and dates back to Ewald \cite{Ewald_lattice_sum}, and also appears in the study of the electronic structure in crystals \cite{PhysRev.92.569}.

In order to circumvent the numerically expensive and inaccurate real space representation in \eqref{eigen_matrix}, we follow an approach introduced by Silveirinha and Fernandes \cite{Silve2005} for the Helmholtz Green function $\mathcal{G}_\mathcal{H}(\vec{r})$. Their main idea is to split the lattice sum into real space contribution and a spectral contribution, using a radial weight function $f(r)$ that approaches $1$ at exponential (or even Gaussian) rate for $r\rightarrow\infty$ and vanishes at the origin $r=0$, thereby splitting $M=Mf(r)+M[1-f(r)]$. The first term is then effectively calculated in real space, whereas the second term converges rapidly in the spectral representation, for which a plane wave expansion yields a summation over reciprocal vectors $\vec{G}$ instead of the real space vectors $\vec{T}$.

This technique can be trivially generalised to the dyadic case exploiting \eqref{Green}. To highlight the importance of this method, we note that the procedure enables us to obtain converged solutions, with relative errors in the eigenvalues at machine precision, for cut-offs as small as $N=5$. For comparison, the naive representation \eqref{eigen_matrix} yields relative errors as large as $10^{-2}$ for $N=100$. Under these circumstances, numerical summation over almost $10^7$ elements per Bloch wave vector makes band structure calculations slow, while still inaccurate and suffering from ringing artefacts. On the contrary, the eigenvalue problem itself is of dimension $12$ in our example and comes at virtually no computational cost.

\section{Hamilton formulation and energy flow \label{app:Hamilton}}
Consider the coupled harmonic oscillator Hamiltonian
\begin{equation}
    H = \frac{1}{2} \left[ \sum_n \left( \bvec{\Pi}_n^2 + \omega_r^2 \bvec{p}_n^2 \right) + \omega_r^2 \sum_{nm} \bvec{p}_n^T I_{nm}^2 \bvec{p}_m \right]
    \label{eq:Ham}
\end{equation}
with the dipole moments $\bvec{p}_n$ taking the role of canonical coordinates, with conjugated momenta $\bvec{\Pi}_n = \frac{d\bvec{p}_n}{dt}$.\footnote{Note that the dipole moments $\vec{p}_n$ in \eqref{dipoles} denote the phasors of monochromatic physical dipole moments $\bvec{p}_n$. Further note that the above Hamiltonian has the odd dimension $(\text{charge}\times\text{length}/\text{time})^2$. This has no direct implication on the following, particularly on the dimensions of the mode velocity. However, scaling the coordinates with $\sqrt{m}/e$, with the effective mass $m$ and the charge $e$ of the microscopic charges, yields correct physical dimensions.} The canonical coordinates and momenta are proportional to the centre of charge with respect to the corresponding sphere centre and the average charge flow, respectively (see also \cite{PhysRevLett.110.106801}).

With the single particle resonance $\omega_r:=\omega_p/\sqrt{3}$ and the interaction matrix given by $I_{nm}:=-\rho^3\mathcal{G}(\vec{r}_n-\vec{r}_m)$ for $n\ne m$ and $I_{nn}=0$, it is straightforward to show that the equations of motion of this Hamiltonian reduce to \eqref{dipoles} in the monochromatic case.

The Hamilton formulation has the advantage, that we can easily derive expressions for the average energy flow of the plasmonic modes. The single particle energy $E_i=\frac{1}{2}\left(\bvec{\Pi}_i^2 + \omega_r^2 \bvec{p}_i^2\right)$ has the time derivative

\begin{equation*}
    \dot{E}_i = -\omega_r^2 \sum_n \bvec{\Pi}_i^T I_{in}\, \bvec{p}_n\text{.}
\end{equation*}

That means that the amount of energy that flows from point $i$ to point $j$ per time is given by $-\omega_r^2 \bvec{\Pi}_i^TI_{ij}\,\bvec{p}_j$. At the same time, $-\omega_r^2\bvec{\Pi}_j^T I_{ji}\,\bvec{p}_i=-\omega_r^2\bvec{p}_i^T I_{ij}\,\bvec{\Pi}_j$ obviously flows from $j$ to $i$. The total power that is transferred from $i$ to $j$ is hence given by the difference of these two contributions (as physically required, it is antisymmetric by construction):
\begin{equation*}
    P_{ij} = \omega_r^2 \left( \bvec{p}_i^T I_{ij}\,\bvec{\Pi}_j - \bvec{\Pi}_i^T I_{ij}\,\bvec{p}_j \right)\text{ .}
\end{equation*}
The symmetric energy flow between $i$ and $j$ can now be defined as
\begin{equation*}
    \vec{S}_{ij} = P_{ij}\left( \vec{r}_j - \vec{r}_i \right)\text{ .}
\end{equation*}
The total energy flow of a system of spheres, averaged over time for the monochromatic field is then given by
\begin{equation}
    \langle\vec{S}\rangle_t = \omega_r^2 \frac{\omega}{4}\sum_{nm} \text{Im}\left\{ \vec{p}_n^\dagger I_{nm}\, \vec{p}_m \right\} \left( \vec{r}_m-\vec{r}_n \right)\text{ ,}
    \label{eq:flow}
\end{equation}
whereas the time average of the total energy is
\begin{align}
    \langle E \rangle_t = &\frac{1}{4} \big[ \left( \omega^2+\omega_r^2 \right)\sum_n |\vec{p}_n|^2 \notag \\
                            & + \omega_r^2\sum_{nm} \text{Re}\left\{ \vec{p}_n^\dagger I_{nm} \vec{p}_m \right\} \big]\text{ .}
    \label{eq:energy}
\end{align}
The mode velocity is then given by
\begin{equation}
    \vec{v} = \frac{\langle \vec{S}\rangle_t}{\langle E \rangle_t}
    \label{eq:velocity}
\end{equation}

For the periodic system, we can replace $\sum_n\mapsto N\sum_\mu$ and $\sum_{nm}\mapsto N\sum_{\mu\nu}$ in \eqref{energy}, if we also replace $I_{nm}\mapsto I_{\mu\nu}:=-\rho^3 M_{\mu\nu}$ and $\vec{p}_n\mapsto\vec{p}_\mu$. The procedure is similar to the one that led to \eqref{eigen_matrix} above. For the flow, we have
\begin{align*}
    &I_{nm}(\vec{r}_m-\vec{r}_n)_i\\
    \mapsto I_{\mu\nu}^{(i)}
        &:= -\rho^3\sum_n \left[ \vec{r}_\mu-\vec{r}_\nu-\vec{T}_n \right]_i \mathcal{G}\left( \vec{r}_\mu-\vec{r}_\nu-\vec{T}_n \right)e^{\imath\vec{k}\cdot\vec{T}_n}\\
        &\phantom{:}= -\rho^3\left( \left[ \vec{r}_\mu-\vec{r}_\nu \right]_i M_{\mu\nu} + \imath\partial_{k_i} M_{\mu\nu} \right)
\end{align*}
so that, in the lossless Drude case, the average energy velocity reduces to the simple expression for the group velocity, as is well known for photonic crystals \cite{JoannopoulosJohnsonWinnMeade:2008}
\begin{equation}
    \vec{v} = \omega_r\,\nabla_{\vec{k}}K_0(\vec{k}) = \nabla_{\vec{k}}\omega(\vec{k})\text{ ,}
    \label{eq:group_velocity}
\end{equation}
where we have used \eqref{eigen} and the Hellmann-Feynman theorem. Note that for any more sophisticated material model $\alpha(k_0)$, the substitution of \eqref{flow} and \eqref{energy} into \eqref{velocity} yields the general form
\begin{widetext}
\begin{equation*}
    \vec{v} = \frac{-2\text{Im}\left\{ \alpha^{-1} \right\} \frac{\sum_\mu \vec{r}_\mu|\vec{p}_\mu|^2}{\sum_\mu |\vec{p}_\mu|^2}-\nabla_{\vec{k}}\left( \text{Re}\left\{ \alpha^{-1} \right\} - \text{Im}\left\{ \alpha^{-1} \right\} \right)}{\left( 1+K_0^2 \right) - \text{Re}\left\{ \alpha^{-1} \right\}} k_0\,c\text{ ,}
\end{equation*}
for which we implicitly assume $\alpha^{-1}=\alpha^{-1}(K_0(\vec{k}))$. This expression does not reduce to the group velocity \eqref{group_velocity}, even for a more general lossless dispersion.
\end{widetext}

\end{document}